\documentclass[preprint, 10pt]{elsarticle}
\pdfoutput=1
\usepackage{graphicx}
\usepackage{subcaption}
\usepackage{amssymb}

\begin{document}
\begin{frontmatter}

\title{DiscretizationNet: A Machine-Learning based solver for Navier-Stokes Equations using Finite Volume Discretization}
\author{Rishikesh Ranade\corref{cor1}\fnref{label1}}
\cortext[cor1]{Corresponding Author.: Email Address: rishikesh.ranade@ansys.com (R.Ranade)}

\author[label2]{Chris Hill}
\author[label1]{Jay Pathak}
\address[label1]{Ansys Inc., Canonsburg, Pennsylvania, USA}
\address[label2]{Ansys Inc., Lebanon, New Hampshire, USA}

\begin{abstract}
Over the last few decades, existing Partial Differential Equation (PDE) solvers have demonstrated a tremendous success in solving complex, non-linear PDEs. Although accurate, these PDE solvers are computationally costly. With the advances in Machine Learning (ML) technologies, there has been a significant increase in the research of using ML to solve PDEs. The goal of this work is to develop an ML-based PDE solver, that couples’ important characteristics of existing PDE solvers with ML technologies. The two solver characteristics that have been adopted in this work are: 1) the use of discretization-based schemes to approximate spatio-temporal partial derivatives and 2) the use of iterative algorithms to solve linearized PDEs in their discrete form. In the presence of highly non-linear, coupled PDE solutions, these strategies can be very important in achieving good accuracy, better stability and faster convergence. Our ML-solver, DiscretizationNet, employs a generative CNN-based encoder-decoder model with PDE variables as both input and output features. During training, the discretization schemes are implemented inside the computational graph to enable faster GPU computation of PDE residuals, which are used to update network weights that result into converged solutions. A novel iterative capability is implemented during the network training to improve the stability and convergence of the ML-solver. The ML-Solver is demonstrated to solve the steady, incompressible Navier-Stokes equations in 3-D for several cases such as, lid-driven cavity, flow past a cylinder and conjugate heat transfer. 
\end{abstract}


%\begin{keywords}
%avc
%\end{keywords}

\end{frontmatter}

\section{Introduction}
The coupling of physics and deep learning to solve problems in the engineering simulation space has drawn interest in recent years. In the context of neural networks, this has been achieved by constraining the network optimization by embedding physical constraints in the loss formulation. These physics-based constraints ensure that the solution space is bounded and obeys physical laws. Although this idea was proposed back in the 90s \citep{lee1990neural, lagaris1998artificial}, it has started to have a big impact in recent times due to rapid advances in computational sciences and deep learning. 
\par Within the context of physics-based deep learning there are two types of methods used to approximate the partial differential equations (PDEs) governing physical processes: data-driven and data-free. The data-driven methods use simulation or experimental data to construct models while enforcing physical laws. These models heavily depend on the fidelity of the data and hence are limited in accuracy and generalizability. On the other hand, data-free methods use neural networks to generate solutions by rigorously constraining the partial differential governing equations through the loss formulation. In this respect, Raissi et al. \cite{raissi2018hidden, raissi2017physics} recently introduced the Physics Informed Neural Network (PINN) framework which approximates the partial derivatives of the solution variables with respect to space and time using automatic differentiation (AD) \citep{baydin2017automatic}. The partial derivatives are used to estimate the PDE losses which are back-propagated to the neural network for weight updates. Raissi et al. \cite{raissi2018hidden, raissi2017physics} demonstrated the PINN methodology to solve a number of 1-D and 2-D partial differential equations (PDEs). A number of studies have used the PINN framework to solve more complex PDEs such as the Navier-Stokes equations. Dwivedi et al. \cite{dwivedi2019distributed} developed a Distributed-PINN to address some of the issues with PINN and demonstrated it to solve Navier-Stokes equation in a lid-driven cavity at low Reynolds numbers. Sun et al. \cite{sun2020surrogate} demonstrated the PINN methodology for surrogate modeling of fluid flow at low Reynolds numbers. Zhu et al. \citep{zhu2019physics} implemented physical constraints on an encoder decoder network in conjunction with flow based conditional generative models for stochastic modeling of fluid flows. Rao et al. \cite{rao2020physics} demonstrated the PINN approach to solve Navier-Stokes equations at low Reynolds numbers for a two-dimensional flow over a cylinder. Very recently, Jin et al. \cite{jin2020nsfnets} proposed the PINN approach for solving Navier-Stokes equations in both laminar and turbulent regimes.
\par The computation of the PDE loss and the choice of network architecture used for network optimization become crucial when solving for highly non-linear, multi-dimensional, stiff, coupled PDEs such as the system of Navier-Stokes equations. The highly non-linear solution space accessed by the Navier-Stokes equations may be challenging to resolve due to the presence of sharp local gradients in a broad computational domain. As a result, the methodology used for computing gradients as well as the approach of network training can be very important in achieving accurate solutions, better stability and faster convergence in the training process.
\par Traditional PDE solver technologies developed over the last few decades have primarily relied on solving discretized formulations of PDEs using methods such as, finite volume, finite element or finite difference. The exact or approximate forms of the linearized discrete equations are used, in combination with linear equation solvers, to improve the solutions iteratively. The discretization method allows access to higher order and advanced numerical approximations for partial derivatives which can be useful in resolving highly non-linear parts of the PDE solution and also add artificial dissipation to improve solver stability. Additionally, these schemes coupled with the iterative solution strategy have proved to be robust in terms of solver stability and convergence. Recently, machine-learning based models have been developed to either learn new discretization schemes from solution data \citep{zhuang2020learned, bar2018data} or to mimic these schemes through novelties in neural network architectures \citep{hsieh2019learning, stevens2020finitenet}. On the other hand, the Ansys suite of software already has access to a large number of advanced discretization schemes that can capture complexities over a wide range of physics. The coupling of these discretization schemes with machine-learning algorithms, along with the iterative solution algorithm, can provide the same benefits in ML-based solvers as observed with traditional solvers.   
\par The main goal of this work is to introduce a new ML-solver, DiscretizationNet, which is a framework that couples solver characteristics with generative networks to solve highly non-linear, multi-dimensional, stiff, coupled PDEs. The solver does not require any training data but generates PDE solutions and simultaneously learns them during the training process. The different finite-volume based numerical schemes are implemented inside the computational graph to enable fast, vectorized operations on GPU and a modified encoder-decoder network architecture is proposed to solve the PDEs in an iterative manner. Finally, the discretization based iterative ML solver is used to solve the steady, incompressible, Navier-Stokes equation in 3-D for a several cases such as, lid-driven cavity flow at a high Reynolds number, flow past a cylinder in laminar regime and conjugate heat transfer. The remainder of the paper is organized as follows. In Section \ref{sec:methodology}, we will introduce the solution methodology adopted in the DiscretizationNet. Subsequently, Section \ref{sec:results} will discuss the numerical results followed with conclusions and future work in Section \ref{sec:conclusion}.  

\section{Solution Methodology}\label{sec:methodology}
In this section, we will discuss the methodology for solving the system of steady, incompressible Navier-Stokes equations using the DiscretizationNet. The system of Navier-Stokes equations consists of the continuity equation and momentum equation for each directional velocity component. The scaled equations described in a vectorized form are given as follows:

\begin{equation}\label{eq:1}
\left. \begin{array}{ll}  
\mbox{\textbf{Continuity Equation:} } \quad\quad\quad \quad\quad\displaystyle\nabla . \textbf{v} = 0\quad\quad \quad\quad\quad\quad\quad\quad\quad\quad\\[8pt]
\mbox{\textbf{Momentum Equation:} } \quad \quad\displaystyle(\textbf{v}.\nabla)\textbf{v} + \nabla \textbf{p} - \frac{1}{Re} \nabla^2 \textbf{\textbf{v}} = 0\quad\quad\quad\quad\\
 \end{array}\right\}
\end{equation}
where \textbf{v} is the scaled velocity vector, $v = (\textit{u, v, w})$, \textbf{p} is the scaled pressure, $\nabla$ is the divergence operator, and $Re$ is the Reynolds number. 

When solving using neural networks, the PDEs described in Eqs. \ref{eq:1} are used to compute the residuals in the loss formulation. Some of reasons that can result in a stiff formulation of this loss term include, complicated geometries, strong coupling of PDE variables in large system of coupled PDEs, presence of multiple domains with different PDE formulations and material properties (such as conjugate heat transfer between fluid and solid domains) and reasonably large Reynolds numbers, where the non-linear, convective component, $(\textbf{v}.\nabla)\textbf{v}$, is dominant. Automatic differentiation (AD) \citep{baydin2017automatic} allows computation of partial derivatives within the computational graph using back-propagation but the stiffness of computed gradients may affect the accuracy, stability and convergence of the neural network training, and may require a large number of training epochs, use of regularization techniques, as well as deep network architectures, which have their own set of problem, e.g. vanishing gradients. At this point, it becomes increasingly important to resolve such PDEs using advanced numerical schemes and develop novel strategies of neural network training. Existing solver methodologies have solved some of these problems and in this work we draw from the vast pool of knowledge to develop our ML-based solver. Next, we introduce the network architecture used in the work followed by the loss formulation and training mechanics. 

\subsection{DiscretizationNet Architecture}

\begin{figure}[hbt!]
  \centerline{\includegraphics[scale=0.35]{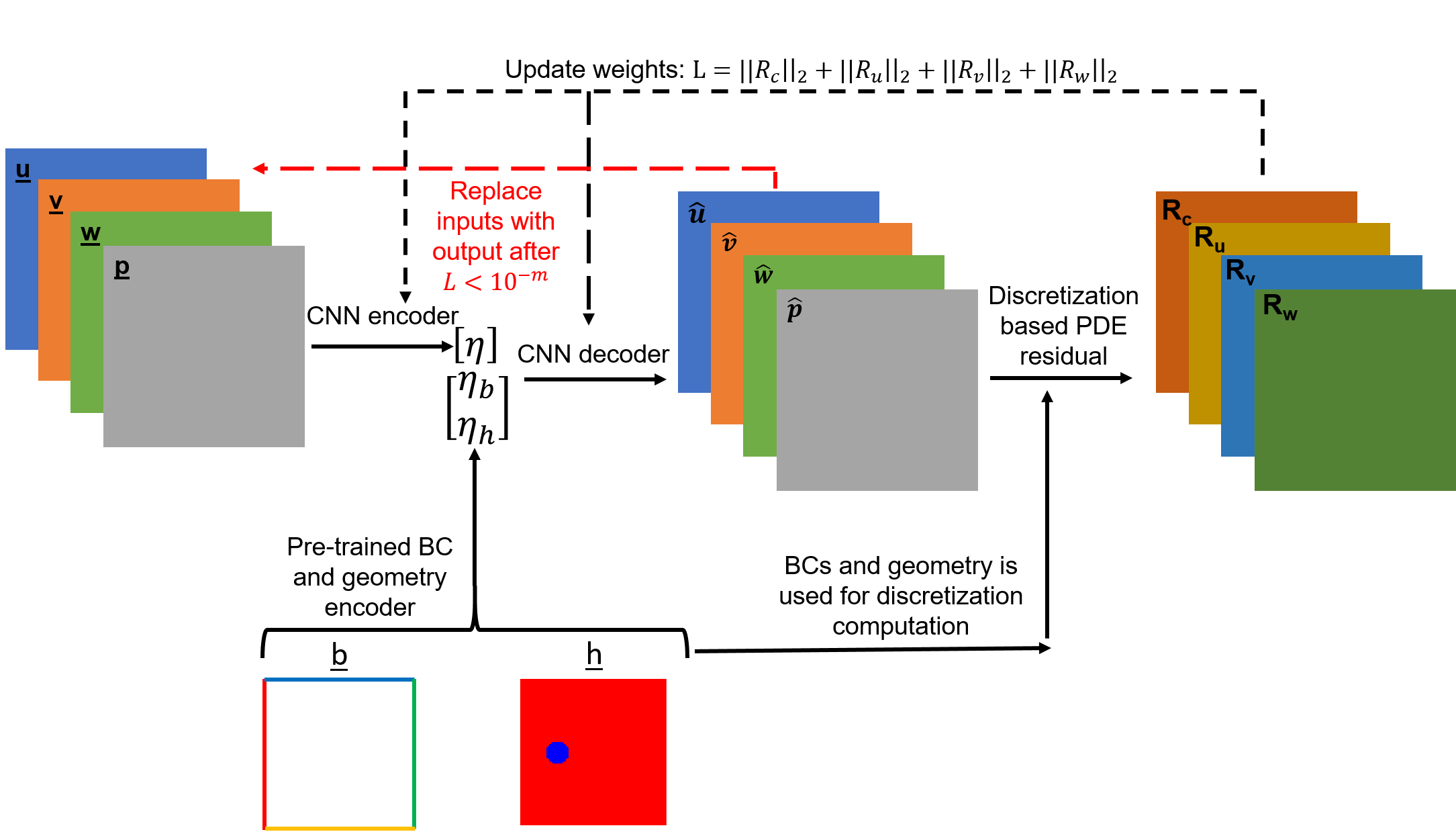}}% Images in 100% size
  \caption{DiscretizationNet for Navier-Stokes solution}
\label{fig:1}
\end{figure}

The network proposed in this work is a generative Convolutional Neural Network (CNN) based encoder-decoder whose input features are flow variables, $v = (\textit{u, v, w, p})$, initialized with random uniform solution fields, boundary condition encoding, $\textit{b}$, and the level set of the geometry, $\textit{h}$. Level sets are real valued functions which depict the geometry such that, regions inside the geometry are flagged with $-1$, the region outside it with $1$ and the regions representing the surface as $0$ \citep{osher1988fronts}. The objective of this network, as shown in, \ref{fig:1}, is to compress the input features $(\underbar u, \underbar v, \underbar w, \underbar p, \underbar b, \underbar h)$ into a lower dimensional space, $(\eta)$, using a convolutional encoder and to decode the latent vector encoding to new solutions, $(\hat u, \hat v, \hat w, \hat p)$, which are closer to actual Navier-Stokes solutions. As can be observed, the boundary conditions as well as the geometry level sets are used to enrich the encoded latent space and then used in the computation of the coupled PDE loss terms, $(R_c, R_u, R_v, R_w)$, corresponding to continuity and momentum equations in each spatial direction. The enrichment of the latent space with geometry and boundary information is crucial as it ensures that the network outputs are conditioned upon them. The geometry and boundary encoders are pre-trained on similar samples and only their weights are required to perform encoding in this network. Encoding geometry and boundary is crucial in this approach, because, in their original form, these representations can be very sparse. This may have an adverse effect on the learning and generalizability of the DiscretizationNet. Additionally, the network proposed here can be used to compute a large batch of solutions at different boundary and geometry conditions in a single training session. As a result, the conditioning of solutions with boundary and geometry conditions enables generalization of PDE solutions for a large set of problems. The purpose of designing the DiscretizationNet as an encoder-decoder network is to obtain a legitimate lower-dimensional encoding of the PDE solution space. The encoded solution space can be useful in developing reduced-order models on top of the DiscretizationNet.

\subsection{Training Mechanics}

It was suggested previously that the input vectors to the encoder-decoder network are randomly initialized solution fields of velocity and pressure. This has two implications, 1) the PDE solution encoding $(\eta)$ does not have a physical meaning and 2) decoding solutions, which are functions of random noise, is a difficult task and may slow down convergence of network and provide poor stability. In order to tackle these challenges, an iterative approach is followed, where the inputs to the network are replaced with the newly generated solutions, every time the PDE residuals reduce by an order of magnitude. At any given point during the training, the solutions generated are dependent on the solutions from a previous iteration and not the initial random solutions used in the beginning. This allows the network to converge from partially converged solutions to fully converged solutions in an iterative fashion and improves stability and convergence as compared to other ML methods in this space. At convergence, when the $L$-2 norm of PDE residuals have dropped to a reasonably low level, the input and output solutions are very similar and effectively turns the network into a conditional autoencoder. It is a conditional autoencoder, since the decoder network is conditioned on the solution encoding as well as the encoding of geometry and boundary. This provides a physical meaning to the reduced dimensional solution latent space, which can now describe flow solutions at given boundary, geometry or flow conditions. Moreover, the PDE solutions are independent of the spatial dimension and depend only on the solutions at previous iterations. This is analogous to how traditional solvers function and provides an opportunity to operate this network under transient conditions. It is important to note that the training is completely data free and the goal here is to generate the solutions by minimizing the PDE residuals and simultaneously learn them into an encoded latent space. 

\subsection{Geometry and Boundary encoder}

\begin{figure}[hbt!]
  \centerline{\includegraphics[scale=0.55]{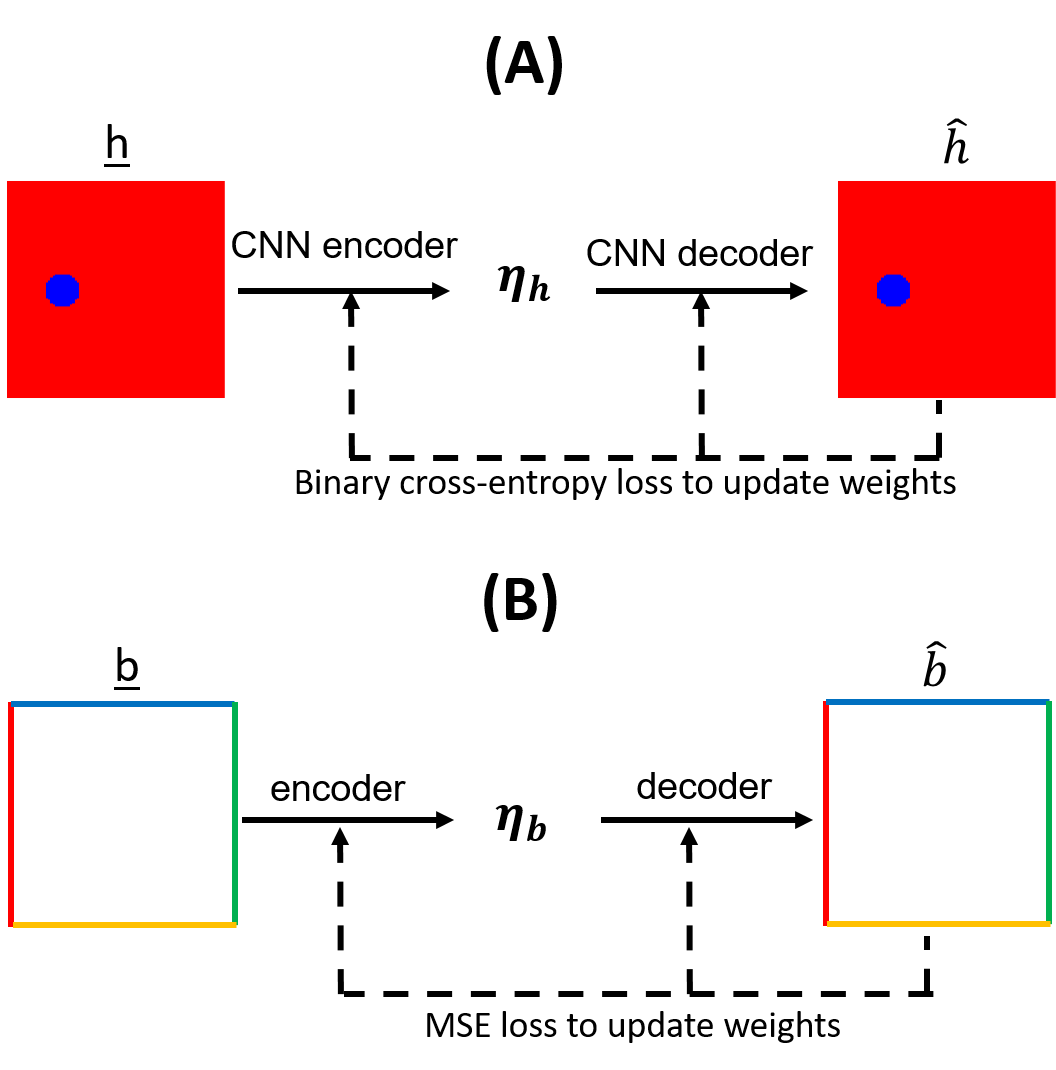}}% Images in 100% size
  \caption{Geometry and boundary autoencoders}
\label{fig:2}
\end{figure}

In this section, we elaborate on the geometry and boundary encoders used in the network architecture in Figure \ref{fig:1}. In this work, we use a modified level set approach to represent the geometry, such that the voxels inside the geometry are represented by \textit{0} and outside by \textit{1}. The gradient of the level set are used to track the voxels activated by the surfaces of the geometry. The level sets for primitive geometries of different shape, size and orientation are learned using a generative encoder-decoder network and represented in a lower-dimensional space, $\eta_h$. A schematic of the geometry autoencoder can be seen in Figure \ref{fig:2}A. In this work, the encoder and decoder networks are CNN-based and a binary cross-entropy loss function is employed to update the weights of the networks. Level sets of different geometries can be generated by parsing through the latent space vector of a trained geometry encoder and used to parameterize the ML-solver.

On the other hand, a separate boundary autoencoder is used to represent different boundary conditions. The boundary encoder is only required if the boundary conditions are spatially or temporally varying. Here, we propose a generative encoder-decoder network to learn the boundary condition encoding but leave the choice of the network architecture open. In scenarios where the boundary condition is constant along the different surfaces, as is in all of the test cases demonstrated in this work, a neural network based boundary condition encoder is not required. Instead, a custom encoding can be constructed and used as the latent vector, $\eta_b$. Flow conditions, such as Reynolds number or Prandtl number can also be perceived as boundary conditions and be added to this latent vector. For example, an encoding of $\eta_b = \left[1, 1, 2, 1, 0.3, 1.2, 0, 3, 40 \right]$ can be perceived as Dirichlet inlet boundary conditions $(1)$ on left, right and bottom surfaces with specified values of $0.3, 1.2$ and $3.0$, respectively, and a Neumann boundary condition $(2)$ on the top surface, where the flux of variable equal zero. The Reynolds number ($40$) or other flow conditions can also be specified in the encoding. A similar choice of boundary condition encoding is employed in this work.  

\subsection{Loss Formulation}

The loss formulation of the network comprises of PDE residual from all the governing equations. Each PDE has its own loss formulation given as follows: 

\begin{equation}\label{eq:2}
\begin{array}{ll}  
 \quad \quad\quad\quad\quad\quad\displaystyle\lambda (\textbf{\textit{W, b}}) = ||\lambda_c||_{\Omega}+||\lambda_u||_{\Omega}+||\lambda_v||_{\Omega}+||\lambda_w||_{\Omega}\quad\quad \quad\quad\quad
 \end{array}
\end{equation}
where $||\lambda_c||_{\Omega}$ is the $L$-2 norm of the continuity residual, $||\lambda_u||_{\Omega}$ is the $L$-2 norm of the \textit{x}-momentum residual, $||\lambda_v||_{\Omega}$ is the $L$-2 norm of the \textit{y}-momentum residual, $||\lambda_w||_{\Omega}$ is the $L$-2 norm of the \textit{z}-momentum residual and $\Omega$ is solution space. Moreover, this network can generate and learn a large set of solutions at different Reynolds numbers. The different solutions simply form a part of the training samples of the encoder-decoder network.

The computation of PDE residuals involves approximation of first and second order spatial gradients. As mentioned earlier, we employ the traditional finite-volume discretization technique to compute the PDE residual loss and moreover, all the numerical schemes are implemented inside the computational graph to enable fast GPU computation. However, this does not limit the use of other discretization schemes such as Finite Element Method (FEM), Discontinuous Galerkin (DG) etc., if higher order elements are required. 

In the finite-volume discretization implemented here, each voxel of the PDE solution is considered as the cell center of an imaginary, finite control volume (CV) as shown in Figure \ref{fig:2}. The volume integrals on the CV are expressed as surface integrals using the Green-Gauss divergence theorem shown below. 
\begin{equation}\label{eq:3}
\int \left( \nabla . v \right) \mathrm{d}V = \sum_j^m v_j n_j A_j
\end{equation}
where, $v$ is a solution variable, $M$ is the number of faces on the CV, $n_j$ is the normal along each face on CV and $A_j$ is the area of each face on CV. Hence, a face-based approach is used to compute gradients across all interior and boundary faces of each control volume in the computation graph. The second order gradients are also computed using Eq. \ref{eq:3}, with the only difference that the solution variable is replaced by its gradient. The convective fluxes are discretized using the first or second-order upwind scheme and the diffusive fluxes are evaluated using a central difference approximation. A second order approximation is used for computing gradients of the pressure field. Since we are dealing with an incompressible formulation of the Navier-Stokes and the discretization is analogous to finite volume discretization on collocated grids, the pressure field is prone to checker-boarding due to a lack of explicit coupling between pressure and velocity and the use of second order numerical schemes. In this work, the pressure-velocity coupling is achieved using the Rhie-Chow interpolation \citep{rhie1983numerical}, which essentially adds a fourth order dissipation of pressure to the continuity equation. The addition of Rhie-Chow flux suffices and more sophisticated schemes such as SIMPLE \citep{patankar1981calculation} are not required but remain an option. In the Rhie-Chow formulation, the velocity at the faces is interpolated as shown in Eq. \ref{eq:10}:
\begin{equation}\label{eq:10}
    u_e = \left( \frac{u_i+u_{i+1}}{2} \right) + \frac{1}{a_p} \left( p_i + \nabla P_i . \bar{r}_i - p_{i+1} - \nabla P_{i+1} \bar{r}_{i+1} \right)     
\end{equation}
where, $u_e$ is the velocity approximation at the \textit{east} face of the control volume, $p$ is the pressure, $\nabla P$ is the gradient of pressure and $a_p$ are the matrix coefficients from the momentum equations.
\begin{figure}[hbt!]
  \centering{\includegraphics[scale=0.27]{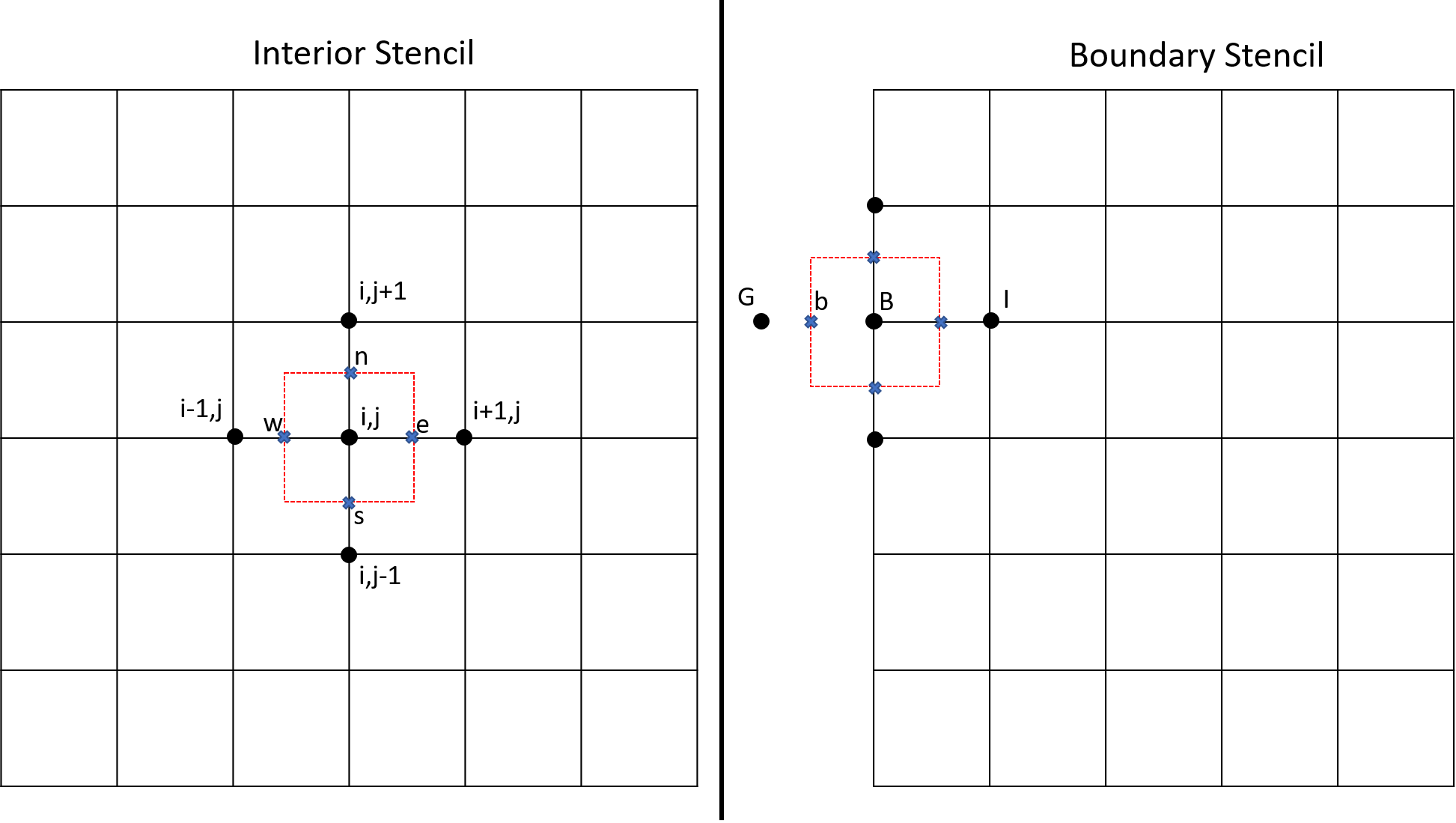}}% Images in 100% size
  \caption{Finite-volume stencil on the interior and boundary pixels in a computational graph}
\label{fig:3}
\end{figure}

The boundary condition treatment is incorporated through the discretization of boundary voxels by enforcing boundary constraints in the flux computation and enforces the order of accuracy at the boundary, as opposed to using one-sided finite difference schemes. For example, the \textit{x}-velocity gradients along a boundary as shown in Figure \ref{fig:2} is represented as follows:
\begin{equation}\label{eq:4}
\left. \begin{array}{ll}  
\quad \quad\quad\quad\displaystyle\left(\frac{\mathrm{d} u}{\mathrm{d} x}\right)_G = 0.5\left(\frac{u_B + u_G}{\Delta x_B}\right) - 0.5\left(\frac{u_B + u_I}{\Delta x_I}\right)\quad\quad\quad\quad\quad\quad\\[16pt]
\quad\quad\quad\quad\quad\quad\quad\quad\quad\quad\quad\quad\displaystyle u_G = 2u_b - u_B \quad\quad\quad\quad\\
\end{array}\right\}
\end{equation}
where, $u_b$ is the specified boundary condition, $u_G$ is an imaginary ghost pixel, $u_B$ is the boundary voxel and $u_I$ is the interior voxel adjacent to the boundary in the direction of the gradient. In the case of unstructured boundaries, such as, cells adjacent to the walls of a cylinder, a stair-step discretization \citep{seo2011sharp} or a cut-cell discretization \citep{tucker2000cartesian} can be implemented. In this work, we have implemented the stair-step discretization, where the boundary conditions at unstructured boundaries are implemented similarly as in Eq. \ref{eq:4}.

The numerical schemes at both interior and boundary voxels are implemented together in the computational graph using vectorized operation for fast computation on GPU. The implementation of the discretization schemes is implemented through custom hidden layers in Keras \cite{chollet2015keras}, where the solution variable tensors as well as boundary and geometry condition tensors are used to compute the PDE residuals of the coupled PDE system at each voxel.  As a result, loss formulation in Eq.\ref{eq:2} contains boundary information and a separate loss term is not needed to model it, as in previous studies \citep{raissi2018hidden, raissi2017physics}, thereby avoiding the need to use strategies for multi-objective optimization. Moreover, the use of discretization techniques to compute loss, allows access to higher order approximations for higher accuracy and advanced numerical schemes such as Rhie-Chow flux \cite{rhie1983numerical} etc. that can enhance stability and convergence of solution and neural network training by providing additional physics-based regularization. 

\subsection{Inference for other geometry and boundary conditions}

The network architecture described in Figure \ref{fig:1} is a generative encoder-decoder network, where, at convergence, the input and output samples are essentially the actual solutions of given PDEs. It may be understood that the model obtained at convergence has learned to encode actual PDE solutions and decode them from the solution latent space combined with geometry and boundary encoding. As a result, the model in its current form cannot be directly used for inferencing solutions at other geometry and boundary conditions, since actual PDE solutions may be required as inputs to the network and these are obviously not available.
\begin{figure}[hbt!]
  \centering{\includegraphics[scale=0.55]{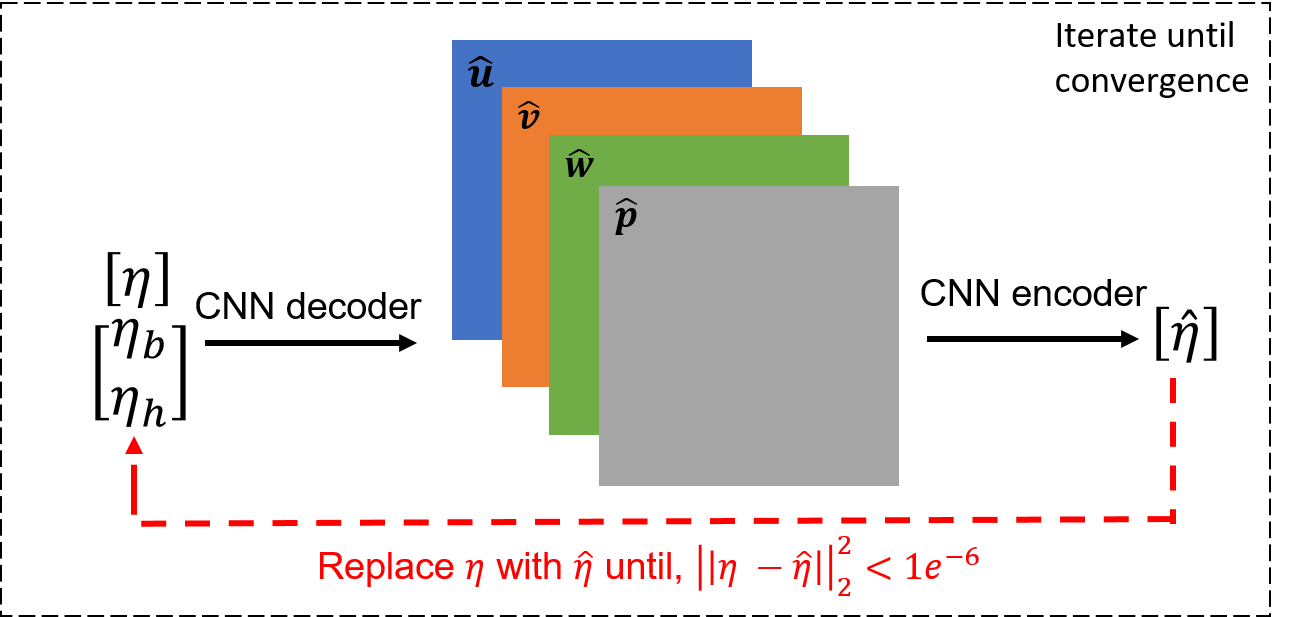}}% Images in 100% size
  \caption{Algorithm for inferencing}
\label{fig:4}
\end{figure}

Here, we propose a novel algorithm that enables inference for other geometry and boundary conditions. A schematic diagram of the algorithm is shown in Figure \ref{fig:4} and the important steps are outlined below. 

\begin{enumerate}
    \item On a given geometry and boundary condition initialized for inference, the geometry and boundary encoding, $\eta_h$ and $\eta_b$ are computed using their respective encoder networks.
    \item Since, the solution encoding, $\eta$, is unknown, it is initialized with a random field drawn from a uniform distribution.
    \item The initial solution encoding combined with the geometry and boundary encoding is passed through the trained weights of the CNN decoder to generate a solution field, $\hat{u}, \hat{v}, \hat{w}, \hat{p}$.
    \item The solution field is encoded to a new solution encoding using the trained weights of the CNN encoder, $\hat{\eta}$.
    \item The new solution encoding, $\hat{\eta}$, replaces the solution encoding of the previous iteration, $\eta$ and steps $(iii)$ and $(iv)$ are repeated until the \textit{L}-2 norm of $\eta-\hat{\eta} < 1e^{-6}$. The geometry and boundary encoding are fixed during the entire process.
    \item At convergence, the PDE solutions at a given geometry and boundary condition are decoded using the most recent $\eta$.
\end{enumerate}

It may be observed that the solution inference happens in the encoded latent vector space and the goal of the iterative procedure is to steer the solution latent vector to a space that is in close proximity to the latent vector spaces observed in the network training. Since the geometry and boundary condition encoding are fixed for a given problem, they provide the necessary constraints for the solution latent vector to iteratively improve itself and generate an accurate PDE solution. The outcome of this algorithm, in terms of generalization, improves with the number of different variations of geometry and boundary conditions adopted during training. Generally, starting from the weights obtained from a well trained model, the inference algorithm converges in fewer than $10$ iterations. Although not a scope of our current work, functioning in the space of latent vectors may provide an opportunity to explore new solution spaces of a given PDE and construct computationally inexpensive reduced order models.

\section{Results}\label{sec:results}
This section provides detailed numerical experiments to demonstrate the ML-solver for several cases of fluid flow such as lid-driven cavity, flow past a cylinder and conjugate heat transfer. The proposed ML-solver is validated against the ANSYS Fluent 19.3 CFD \citep{fluent} solver for solving the incompressible, steady Navier-Stokes equation for these different cases.

\subsection{Lid-driven cavity flow}
In this experiment, the 3-D lid-driven cavity problem is solved at three different Reynolds numbers (Re), $1000$, $2500$ and $5000$. In 3-D, this problem is known to become unsteady for Re numbers above $3000$ \citep{gelfgat2019linear} and may cause convergence issues in CFD solvers for steady, incompressible Navier-Stokes equation. Such issues are not experienced with the ML-Solver, which is able to solve up to Reynolds numbers of $10,000$ and even higher, due to inherent regularization in the ML techniques. As a result, the Reynolds numbers of $1000$ and $2500$ are chosen to obtain fair comparisons with traditional CFD solvers, since they fall in the steady regime and yet exhibit substantial non-linearity. Streamline plots at Reynolds number of $5000$ are presented to demonstrate the ability of the ML-solver to operate at higher Reynolds numbers. Moreover, it is important to note that these solutions are different training samples to a single neural network and hence, they are all learned and generated simultaneously.

\begin{figure}[hbt!]
  \centerline{\includegraphics[scale=1.0]{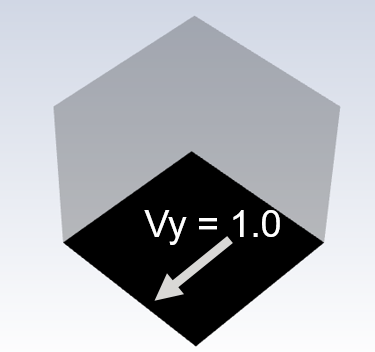}}% Images in 100% size
  \caption{Description of computational domain for lid-driven cavity flow}
\label{fig:5}
\end{figure}
A schematic diagram of the problem is shown in Figure \ref{fig:5}. The boundary conditions are applied such that the bottom wall is moving in the \textit{y}-direction with a specified velocity. The boundary velocity and domain dimensions are constant and the kinematic viscosity is tuned with respect to the test Reynolds numbers. The computational domain has a resolution of $64^3$. The implementation of the ML algorithm is carried out in Keras \citep{chollet2015keras}, which provides a lot of flexibility in creating custom hidden layers that are useful in developing the discretization schemes inside computation graphs, as discussed before. The CNN encoder and decoder for all solution variables have 3 layers each with 64 filters in each layer and tangent-sigmoid activation function. The training is carried out using an Adam optimizer with an initial learning rate of $1e^{-3}$. A learning rate scheduler is used to tune the learning rate depending on the network loss. Since, we are using an iterative approach where the input samples are replaced with the output, the learning rate is reinitialized to $1e^{-3}$, every time the iterative operation is implemented. For all test cases, about $3$x$10^4$ training iterations are carried out on a NVIDIA Tesla V100 SXM2 GPU.

\begin{figure}[hbt!]
  \centerline{\includegraphics[scale=0.28,trim=4 4 4 4,clip]{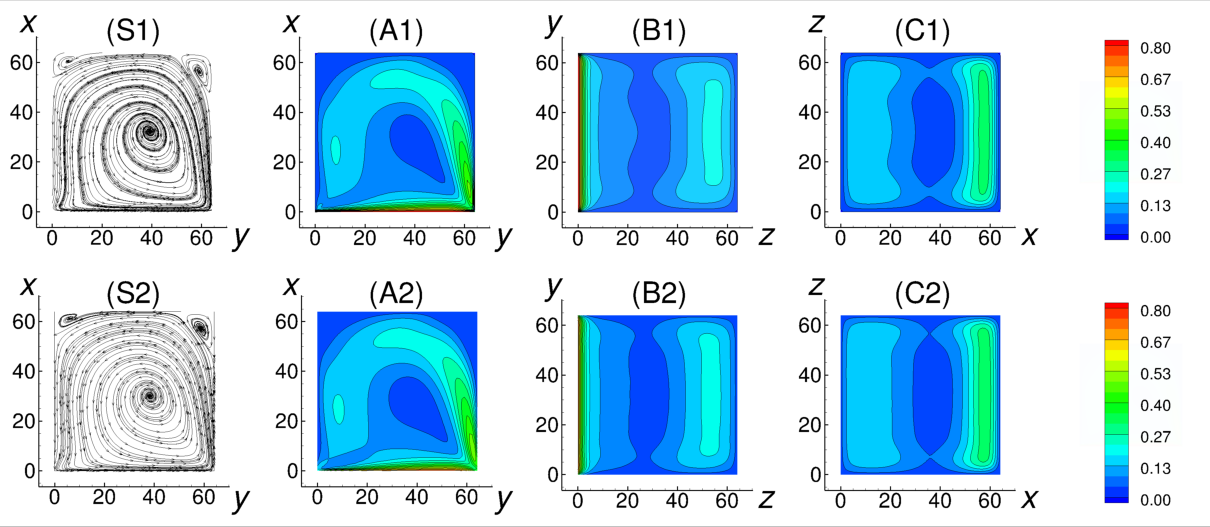}}% Images in 100% size
  \caption{Comparison of velocity magnitude in a Lid-driven cavity at Re = 1000 between ANSYS Fluent(1) and ML-Solver(2) on planes A) \textit{x}-\textit{y}, B) \textit{x}-\textit{z} and C) \textit{y}-\textit{z} cut along the center of the domain}
\label{fig:6}
\end{figure}

\begin{figure}[hbt!]
  \centerline{\includegraphics[scale=0.28,trim=4 4 4 4,clip]{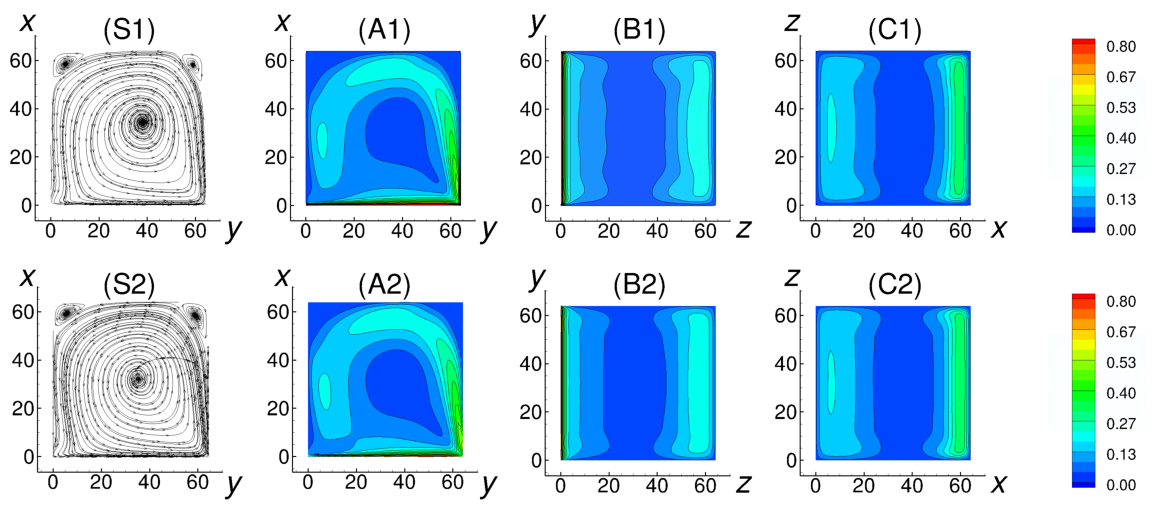}}% Images in 100% size
  \caption{Comparison of velocity magnitude in a Lid-driven cavity at Re = 2500 between ANSYS Fluent(1) and ML-Solver(2) on planes A) \textit{x}-\textit{y}, B) \textit{x}-\textit{z} and C) \textit{y}-\textit{z} cut along the center of the domain}
\label{fig:7}
\end{figure}

The predicted velocity magnitude for both Reynolds numbers are shown in Figures \ref{fig:6} and \ref{fig:7}. The results from the ML-Solver are compared with results generated from ANSYS Fluent R19.3 \citep{fluent}. The discretization schemes used are consistent in both cases. The comparisons are carried out on planes cut through the center of the domain along all three dimensions (A, B and C) as well as a stream line plot along \textit{y}-\textit{z} plane through the center of the domain (S). The contour plots depict the velocity magnitude while the streamlines are plotted for the velocity in the \textit{y}-direction. It may be observed from contour plots that the predictions of the ML-solver match reasonably well with those from ANSYS Fluent. The streamline plots show that the global flow features such as the positions of the primary and the secondary vortical structures are captured well.
\begin{figure}[hbt!]
  \centerline{\includegraphics[scale=0.28,trim=4 4 4 4,clip]{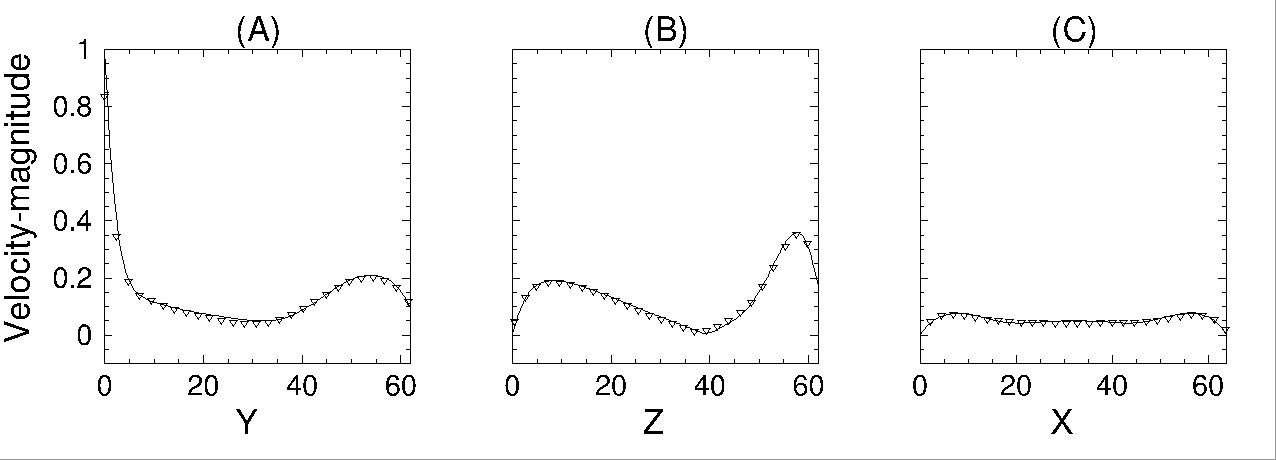}}% Images in 100% size
  \caption{Comparison of velocity magnitude in a Lid-driven cavity at Re = 1000 between ANSYS Fluent(symbol) and ML-Solver(solid) on centerline along A) \textit{y}, B) \textit{z} and C) \textit{x}}
\label{fig:8}
\end{figure}
\begin{figure}[hbt!]
  \centerline{\includegraphics[scale=0.28,trim=4 4 4 4,clip]{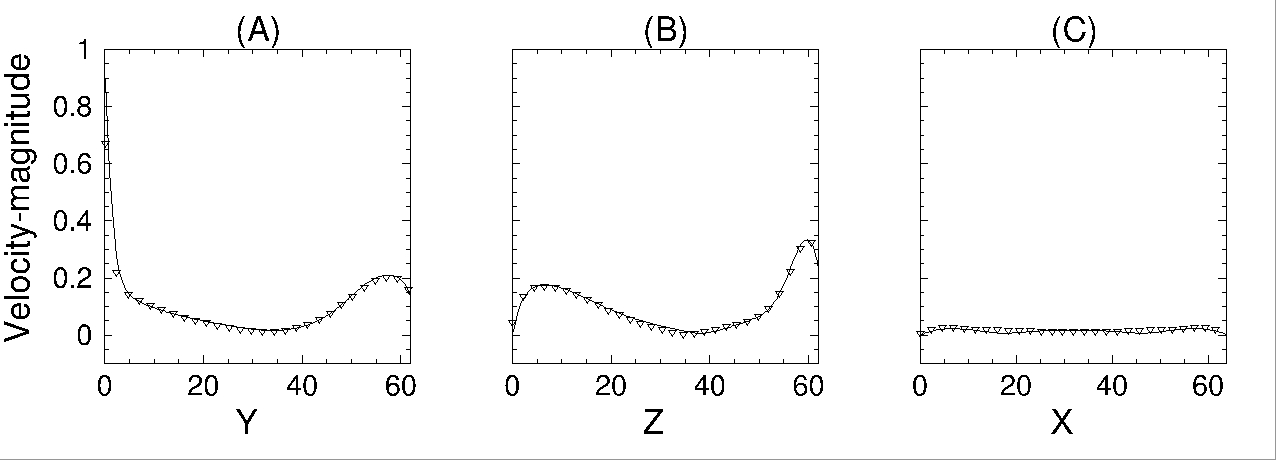}}% Images in 100% size
  \caption{Comparison of velocity magnitude in a Lid-driven cavity at Re = 2500 between ANSYS Fluent(symbol) and ML-Solver(solid) on centerline along A) \textit{y}, B) \textit{z} and C) \textit{x}}
\label{fig:9}
\end{figure}

\begin{figure}[hbt!]
  \centerline{\includegraphics[scale=0.28,trim=4 4 4 4,clip]{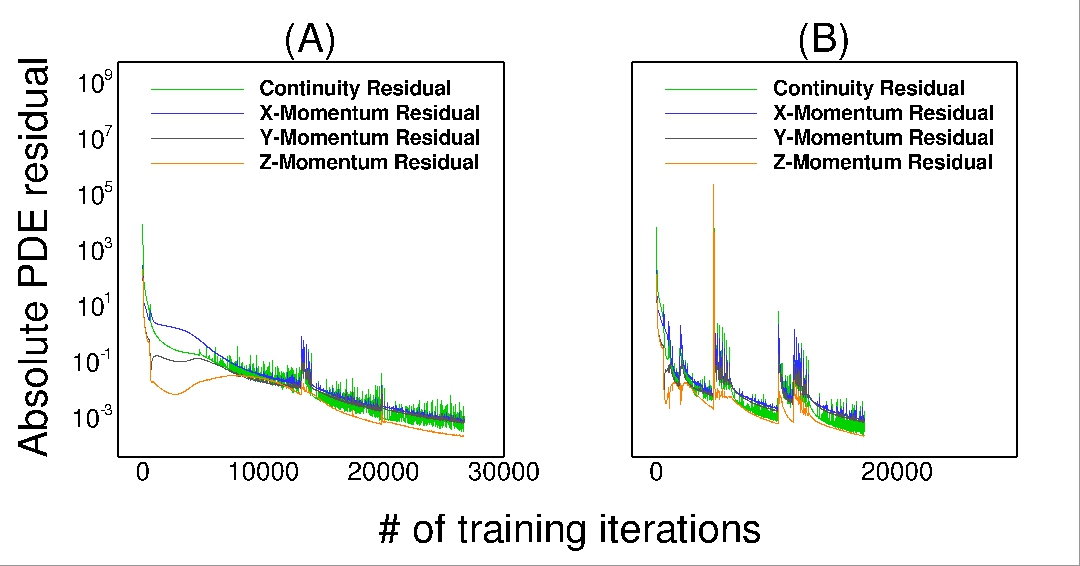}}% Images in 100% size
  \caption{Comparison of residual convergence, A) without using iterative strategy  and  B) using iterative strategy}
\label{fig:10}
\end{figure}
Although, contour and streamline plots provide a good qualitative comparison of the flow structure, the line plots along the centerline of domain provide a more quantitative comparison between the two solutions. In Figures \ref{fig:8} and \ref{fig:9}, the velocity magnitude is plotted on lines passing through the center of the domain along \textit{x}, \textit{y} and \textit{z} directions. It may be observed that the results agree well and the relative errors are less than 1 \% at each point, attributing to the good accuracy of the ML-solver. 

Figure \ref{fig:10} shows a comparison of the convergence of absolute residuals of continuity, \textit{x}-momentum, \textit{y}-momentum and \textit{z}-momentum equations for training with and without the iterative procedure. It may be observed that the convergence of solutions to an absolute residual of $1e^{-3}$ is faster when the iterative procedure is implemented. Spurious peaks may be observed in the convergence plots with iterative procedure. Since the network weights are tuned for a given input vector, the replacement of inputs with outputs causes the PDE residuals to momentarily jump. It is important to note that this behavior is only observed in the beginning, when the output solutions are significantly different from the input vectors. As the PDE residuals reduce and the solutions get closer to convergence, the inputs and outputs have fewer differences and any replacements made after this point results in a smoother convergence. Additionally, it is important to note that in both cases, the total number of training epochs are less than $3$x$10^4$ and each training epoch requires a computation time of about $1$ second. The computation of discretization in the computation graph does not add significantly to the training cost and in fact improves training  stability, thus resulting in fast convergence.
\begin{figure}[hbt!]
  \centerline{\includegraphics[scale=0.29,trim=4 4 4 4,clip]{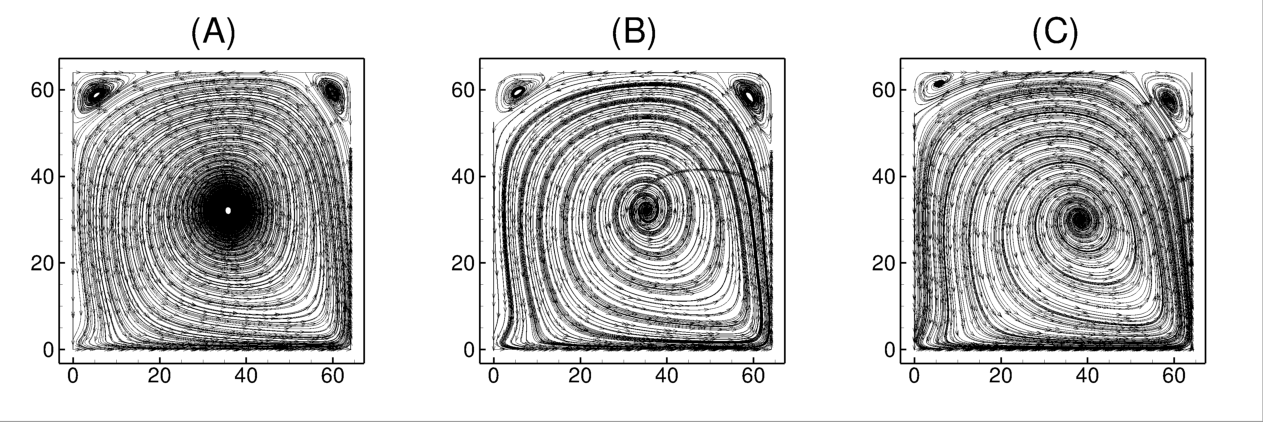}}% Images in 100% size
  \caption{Streamline plot comparisons on the center \textit{x}-\textit{y} in a Lid-driven cavity at Re = A) 5000, B) 2500 and C) 1000}
\label{fig:11}
\end{figure}

Next, we present a comparison of streamline plots obtained from the ML-Solver at $Re = 5000$ in Figure \ref{fig:11} with other Reynolds numbers. The streamlines are plotted along plane cut through the center of the domain along \textit{z} direction. The ML-solver provides convergent solutions for the steady, incompressible Navier-Stokes equation at $Re = 5000$ and even beyond, as a result of the added physics-based regularization, but solutions from other PDE solvers do not converge easily for Reynolds numbers beyond 3000, hence those comparisons are not provided. It may be observed from Figure \ref{fig:11} that an increase in Reynolds number results in clear differences in the size and position of the primary and secondary vortical structures, attributing to the ability of the ML-Solver to capture solutions at higher Reynolds numbers. 

The validation of the solver for the lid-driven cavity case at reasonably high Reynolds numbers shows that the ML-solver can generate accurate non-linear flow solutions as well as result in better stability and fast training in $3-$dimensional scenarios. With that being said, lid-driven cavity is a relatively simpler case where the geometry is not complicated, there is minimal interaction of fluid and solid domains, and the only source of stiffness in the PDEs results from the non-linearity in the solution field due to a high Reynolds number. In the following sections, we validate the ML-solver for cases involving complicated geometries and non-trivial fluid-solid interactions.      

\subsection{Laminar flow past a cylinder}

\begin{figure}[hbt!]
  \centerline{\includegraphics[scale=0.8]{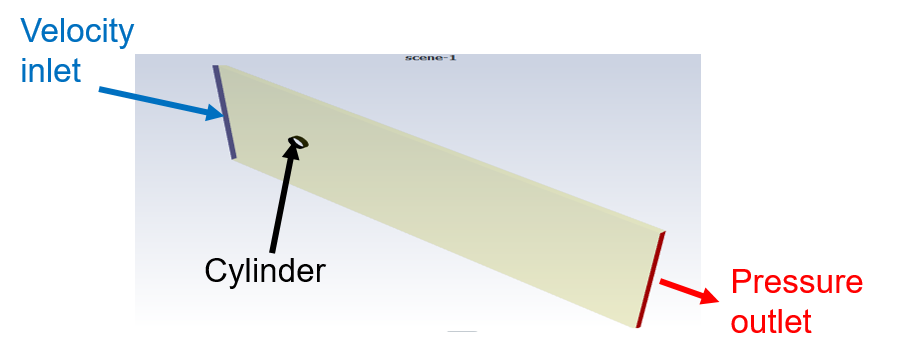}}% Images in 100% size
  \caption{Description of computational domain for flow past a cylinder}
\label{fig:12}
\end{figure}

In this section, we validate the performance of the ML-solver in solving the 3-D, steady, incompressible Navier-Stokes equations for flow past a cylinder at $3$ different Reynolds numbers in the steady regime, $Re = 10, 20, 40$. A schematic diagram of the problem is shown in Figure \ref{fig:12}. The computational domain extends to about $30D$ in the $x$-direction and $10D$ in the $y$ and $z$ directions, where $D$ is the diameter of the cylinder. The left boundary of the computational domain is specified as the velocity inlet, while the right boundary is the pressure outlet. All the other boundaries, perpendicular to the cylinder, are specified as symmetric. The training is carried out using the same network architecture and procedure as described in the previous section. The network is trained to generate solutions at the previously stated Reynolds numbers as well as $5$ different velocity inlet conditions for each Reynolds number given as, $ 0.2, 0.4, 0.6, 0.8, 1.0$.  

\begin{figure}[hbt!]
  \centerline{\includegraphics[scale=0.25,trim=4 4 4 4,clip]{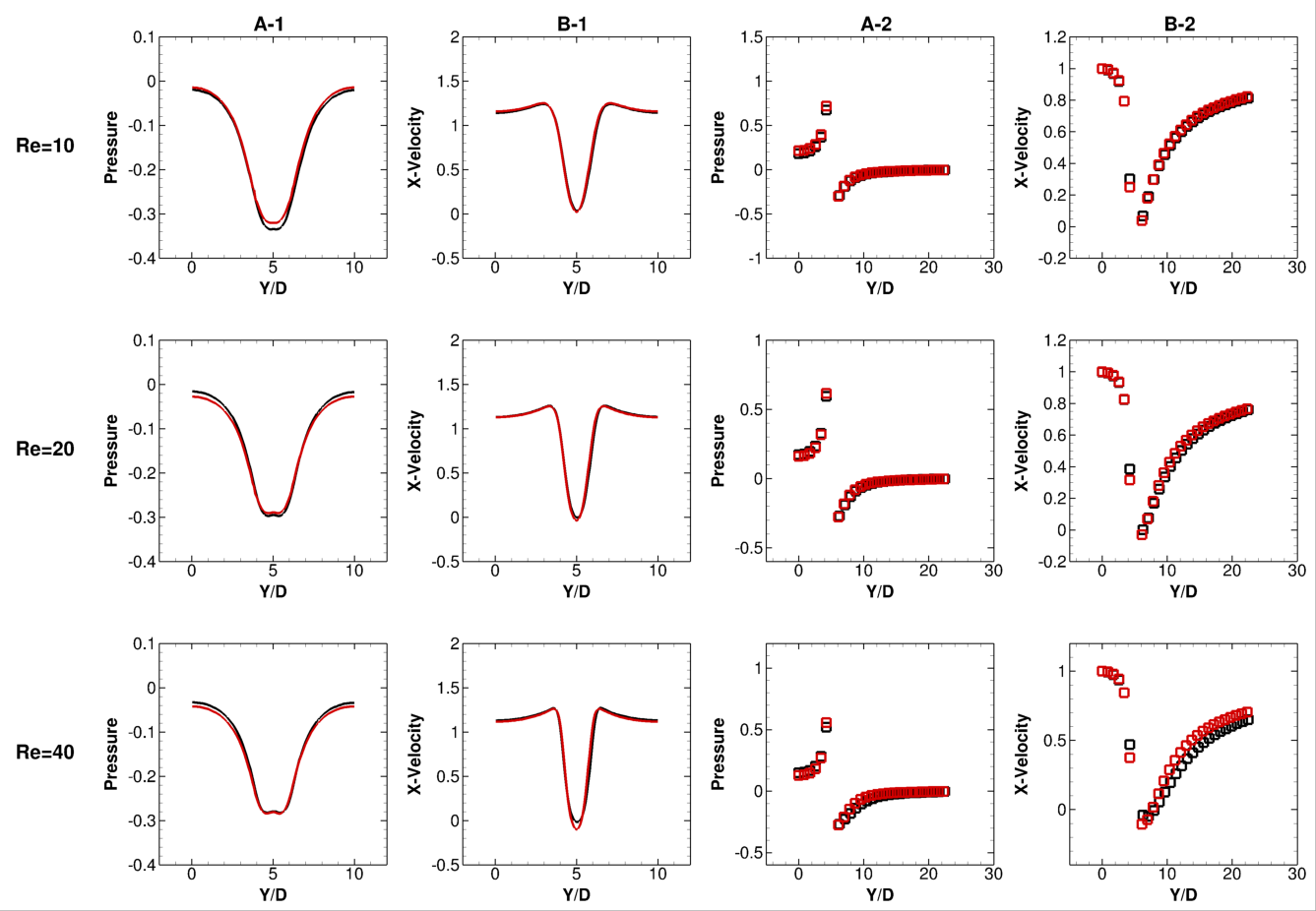}}% Images in 100% size
  \caption{Comparison between ML-solver (red) and Ansys Fluent (black) for $x$-velocity (A) and pressure (B) at different Reynolds numbers plotted at, 1) $x/d=6$ and 2) center line at inlet velocity of 1.0}
\label{fig:14}
\end{figure}

\begin{figure}[hbt!]
  \centerline{\includegraphics[scale=0.25,trim=4 4 4 4,clip]{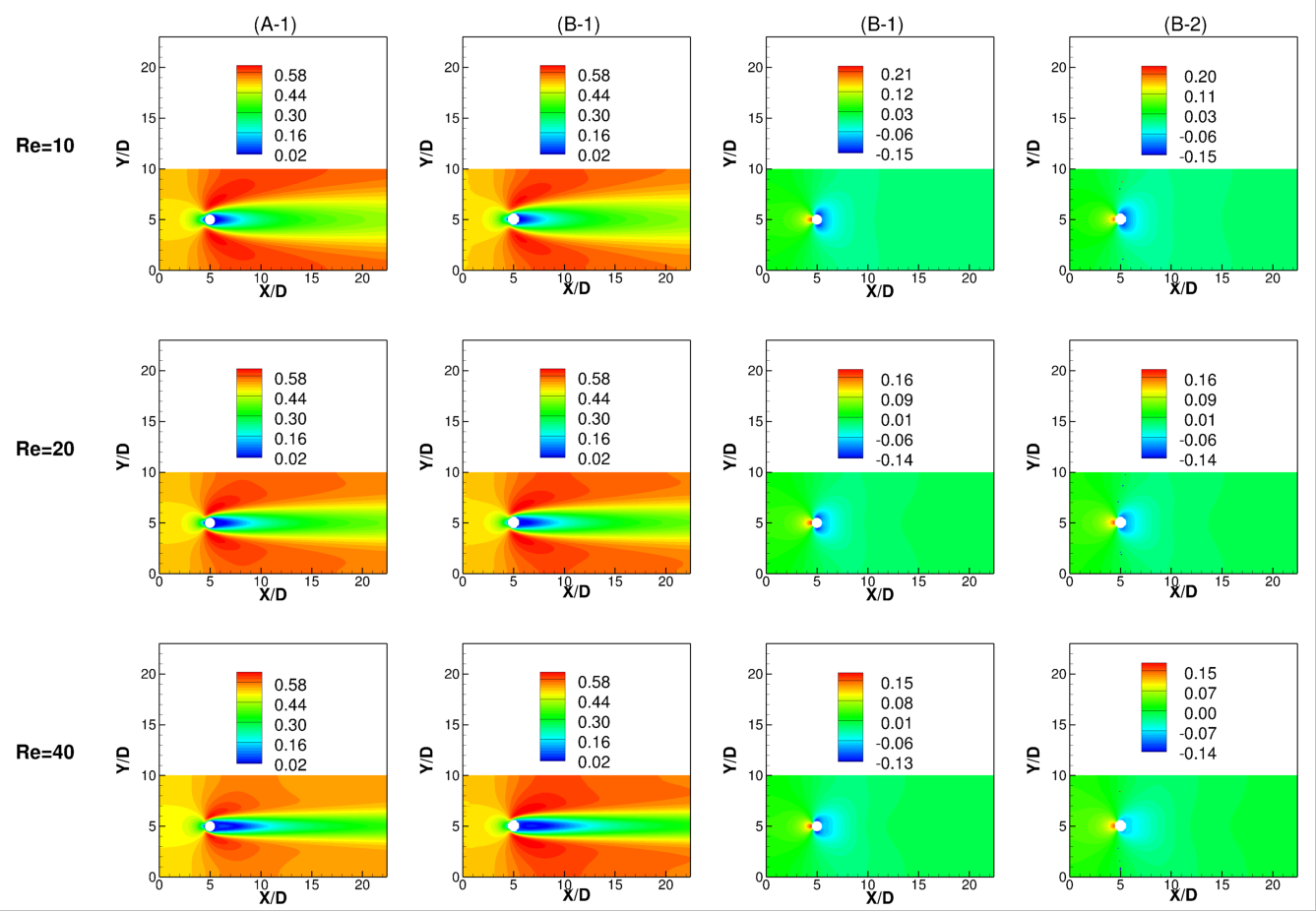}}% Images in 100% size
  \caption{Comparison of velocity magnitude (A) and pressure (B) at different Reynolds numbers between ML-Solver (1) and Ansys Fluent (2) at inlet velocity of 0.5}
\label{fig:15}
\end{figure}

The predicted velocity magnitude as well as pressure for a solution at a velocity inlet boundary condition of $1.0$, generated during training, are presented in Figure \ref{fig:14} for the three Reynolds numbers. It may be observed that the solutions are in great agreement with the solutions obtained from Ansys Fluent 19.3 \cite{fluent} and the flow structures on the upstream and downstream of the cylinder are captured well. Next, we compare line plots of pressure and $x$-velocity at two different locations, $x/d=6$ and center, $y/d=5$ for the same case, as in Figure \ref{fig:14}. The line plot comparisons, presented in Figure \ref{fig:15}, show that the solutions obtained from the ML-Solver are less than $5$\% of solutions obtained from Ansys Fluent \citep{fluent}. Any differences in the solutions can be attributed to the discretization error near the curved surfaces of the cylinder. As mentioned earlier, the ML-solver employs a stair-step discretization to capture the cylinder surface and that can be dissipative, if the mesh is not fine enough. On the other hand, Ansys Fluent uses an unstructured grid discretization, which provides a better representation of the cylinder surface and thus, results in slightly more accurate solutions, as can be observed from the line plots in  Figure \ref{fig:15}. This issue can be circumvented by adding a cut-cell discretization capability \citep{tucker2000cartesian} or an unstructured grid discretization in the ML-Solver and couple it with graph CNNs \citep{kipf2016semi} or Mesh-based CNNs \citep{hanocka2019meshcnn} to perform convolution and pooling operations on unstructured domains.
\begin{figure}[hbt!]
  \centerline{\includegraphics[scale=0.25,trim=4 4 4 4,clip]{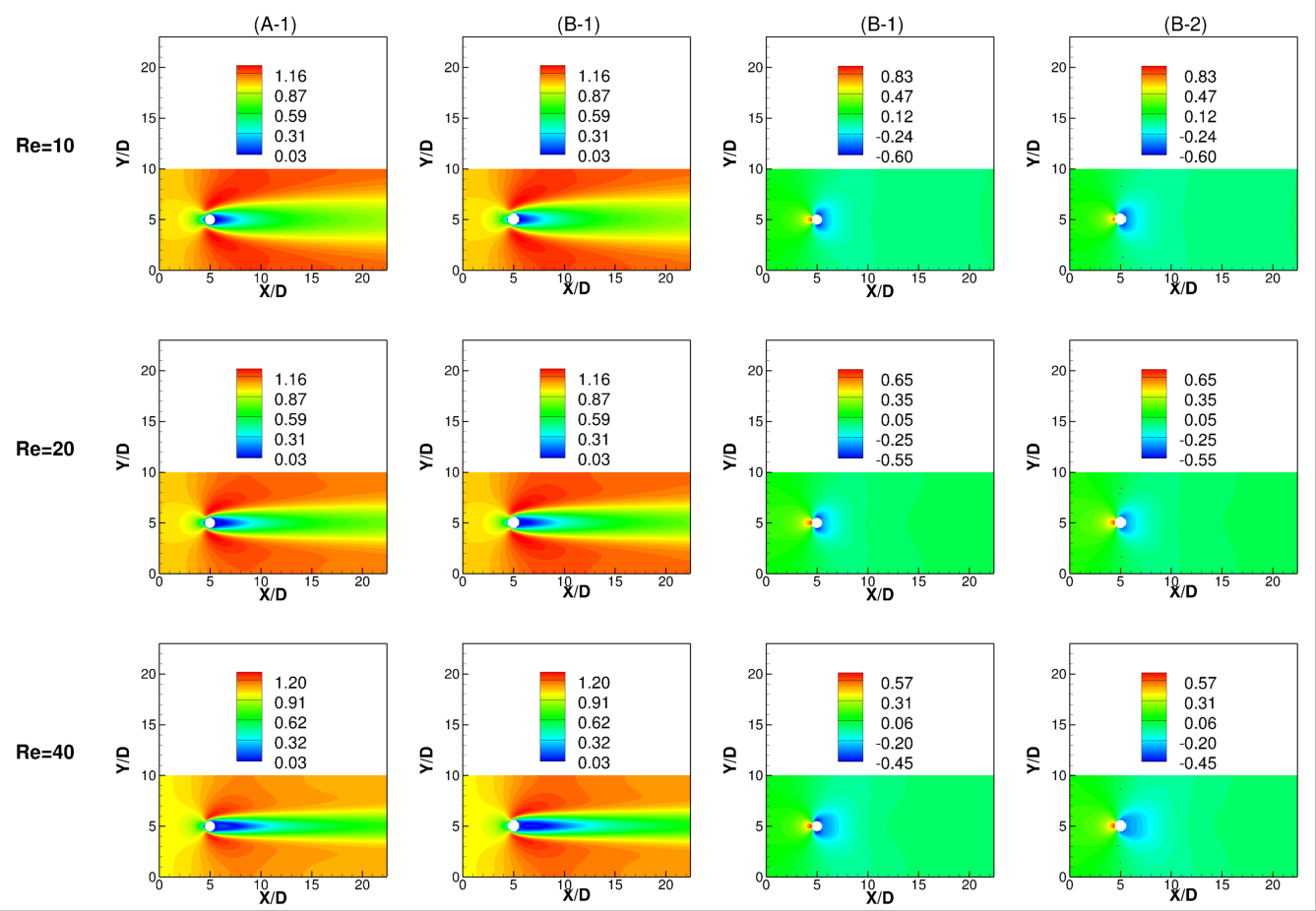}}% Images in 100% size
  \caption{Comparison of velocity magnitude (A) and pressure (B) at different Reynolds numbers between ML-Solver (1) and Ansys Fluent (2) at inlet velocity of 1.0}
\label{fig:13}
\end{figure}

Next, we use the inferencing algorithm described previously in Figure \ref{fig:4} to evaluate the solution at a test inlet velocity condition of $0.5$ for all three Reynolds numbers. The inlet velocity used for testing was not generated or learned during network training. The purpose here is to demonstrate the adequacy and validate the inferencing algorithm. The predictions at the test velocity are shown in Figure \ref{fig:15}. It may be observed that the predictions of generalization algorithm match well with the solutions obtained from Ansys Fluent \citep{fluent}. 

\subsection{Conjugate heat transfer}

\begin{figure}[hbt!]
  \centerline{
  \includegraphics[scale=0.40]{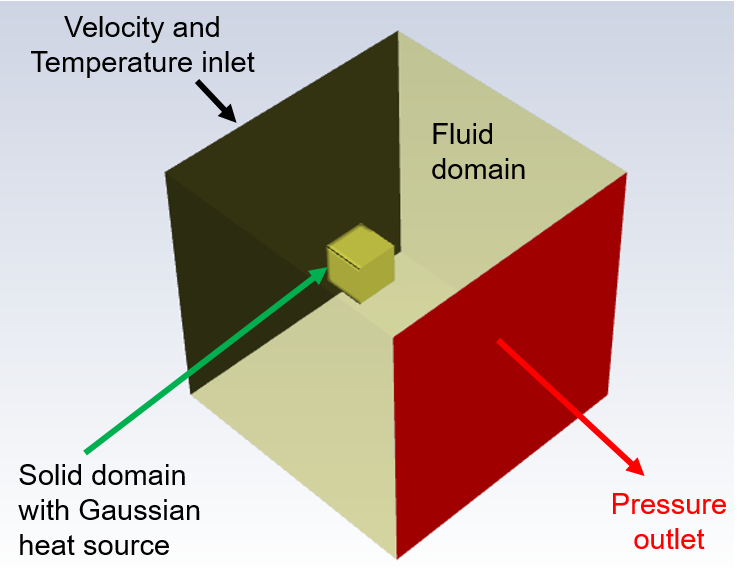}% Images in 100% size
  \includegraphics[scale=0.45,trim=4 4 4 4,clip]{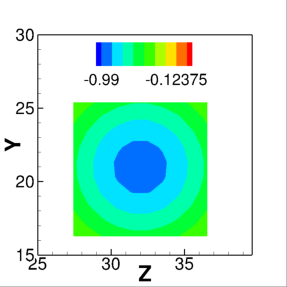}}% Images in 100% size
  \caption{Description of computational domain for conjugate heat transfer (left) and a contour of Gaussian power map on a center-plane along Y-Z (right)}
\label{fig:16}
\end{figure}

Finally, we demonstrate the ML-solver to solve a conjugate heat transfer problem in a laminar flow setting ($Re = 10$) and compare the results with Ansys Fluent \citep{fluent}. A schematic of the problem description in provided in Figure \ref{fig:16}. The computational domain consists of two parts, a fluid and a solid domain. The fluid domain extends $64$ m in each direction with velocity and temperature inlet specified on the left boundary and pressure outlet on the right. All the other surfaces are specified as symmetry. On the other hand, the solid domain is a smaller cube, and is placed inside the fluid domain. Seven different side lengths of solid cubes ranging from, $8$ m and $20$ m, are considered during training. The thermal diffusivity in the solid is assumed to be 2 times as that of the fluid. A Gaussian heat source, depicted in Figure \ref{fig:16}, is described at the center of the solid domain as shown in Eq. \ref{eq:6}. 
\begin{equation}\label{eq:6}
P\left( \bar{x} \right) = 0.05 * \left( \frac{1}{\sqrt{(2 \pi)^3 det(\Sigma)}} exp\left( -\frac{1}{2} (\bar{x}-\mu)^T \Sigma^{-1} (\bar{x}-\mu) \right) \right)
\end{equation}
where, $\mu$ is the mean and $\Sigma$ is the covariance matrix, where the variance in each direction equal to $0.15$.

The network architecture and procedure used during training is similar to previous experiments. The only difference is in the loss formulation, where different equations are solved in the fluid and solid domains. In the fluid domain, we solve the Navier-Stokes equation, Eq. \ref{eq:1}, and Energy equation, Eq. \ref{eq:5}, while the Heat Conduction equation is solved in the solid domain, Eq. \ref{eq:5}. A one-way coupling is used at the interface between the fluid and solid domains. The Energy and Heat equations are shown in Eq. \ref{eq:5}.
\begin{equation}\label{eq:5}
\left. \begin{array}{ll}  
\mbox{\textbf{Heat Equation:} } \quad\quad \quad\quad \quad\quad\displaystyle\nabla .\left( \alpha\nabla\textbf{T}\right)-P = 0\quad\quad \quad\quad\quad\quad\quad\quad\\[8pt]
\mbox{\textbf{Energy Equation:} } \quad\quad\quad\quad\displaystyle(\textbf{v}.\nabla)\textbf{T} - \nabla .\left( \alpha\nabla\textbf{T}\right) = 0\quad\quad\quad\quad\quad\\
 \end{array}\right\}
\end{equation}
where \textbf{v} is the velocity vector, $v = (\textit{u, v, w})$, \textbf{T} is the normalized temperature based on inlet temperature, $\nabla$ is the divergence operator, $P$ is the heat source, and $\alpha$ is the thermal diffusivity. The Navier-Stokes equation and energy equation discretization's are employed in the fluid domain while the Heat equation discretization with power source is used in the solid domain. The solution for an additional PDE for energy as well as the contrasting properties and the coupling between the solid and fluid domains results in additional challenges for the ML-solver as compared to the case of flow past a cylinder, presented previously.  

\begin{figure}[hbt!]
  \centerline{\includegraphics[scale=0.35,trim=4 4 4 4,clip]{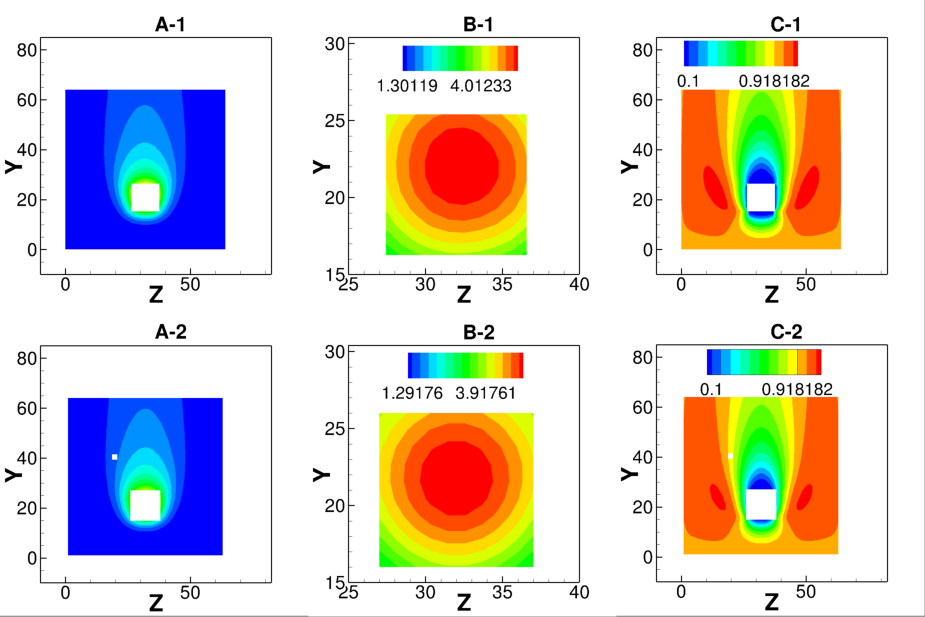}}% Images in 100% size
  \caption{Comparison of normalized temperature in fluid domain (A), solid domain (B) and \textit{y}-velocity (C) between ML-solver (1) and Ansys Fluent (2) for a solid cube of size $10$m}
\label{fig:17}
\end{figure}

\begin{figure}[hbt!]
  \centerline{\includegraphics[scale=0.37,trim=4 4 4 4,clip]{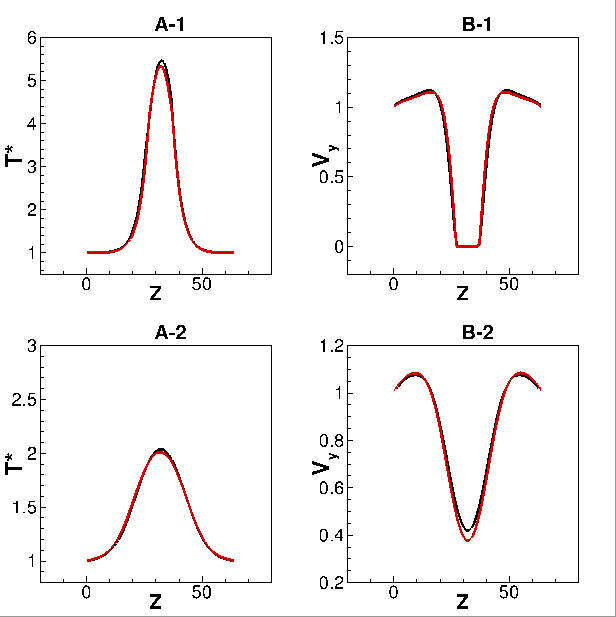}}% Images in 100% size
  \caption{Comparison of normalized temperature (A) and \textit{y}-velocity (B) between ML-solver (red) and Ansys Fluent (black) at \textit{y} = 21 (1) and \textit{y} = 40 (2) for a solid cube of size $10$m}
\label{fig:18}
\end{figure}

Next, we compare the predicted normalized temperature and velocity profile generated by the ML-solver during training with solutions from Ansys Fluent \citep{fluent} for a $10$m long solid cube. The contour plots in Figure \ref{fig:17} show the temperature profile in the fluid and the solid domain, as well as the velocity profile in the fluid domain. It may be observed that the peak temperature occurs inside the solid domain due to the presence of a heat source and the temperature from the solid dissipates into the fluid. The direction of dissipation and the magnitude of peak temperature are affected by the velocity profile and the flow Reynolds number. It may be observed from Figure \ref{fig:17}, that the solutions from ML-solver match reasonably well with the Ansys Fluent 19.3 \citep{fluent} solutions.  
Figure \ref{fig:18} shows line plot comparisons of normalized temperature and $y$-velocity between the ML-solver and Ansys Fluent \citep{fluent} at $y=21$ and $y=40$. The ML-solver results are withing 1\% of the Fluent solutions in both fluid and solid domains, thus validating the accuracy of the ML-solver. 

\begin{figure}[hbt!]
  \centerline{\includegraphics[scale=0.35,trim=4 4 4 4,clip]{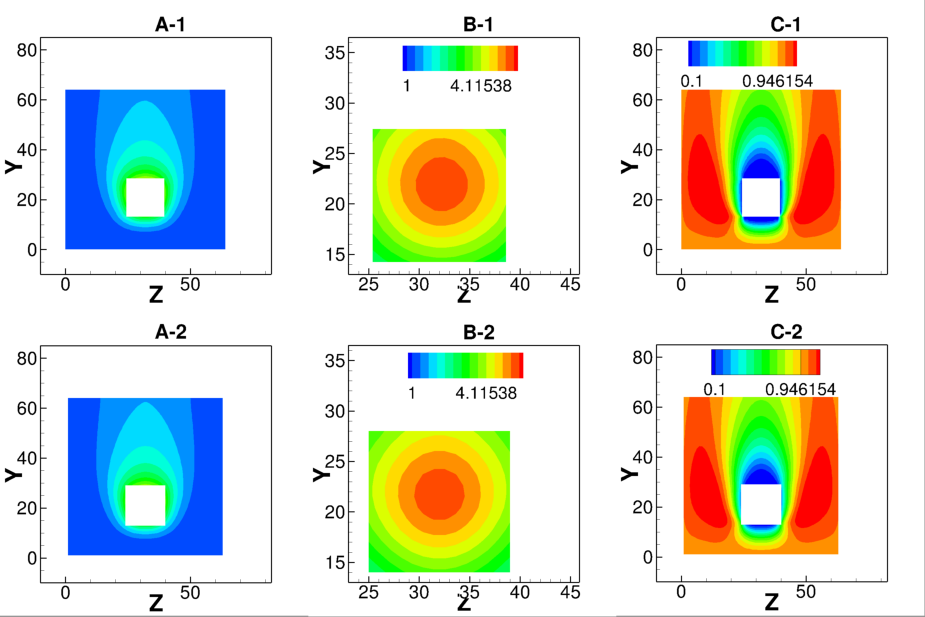}}% Images in 100% size
  \caption{Comparison of normalized temperature in fluid domain (A), solid domain (B) and \textit{y}-velocity (C) between ML-solver (1) and Ansys Fluent (2) for a solid cube of size $13$m}
\label{fig:19}
\end{figure}

\begin{figure}[hbt!]
  \centerline{\includegraphics[scale=0.35,trim=4 4 4 4,clip]{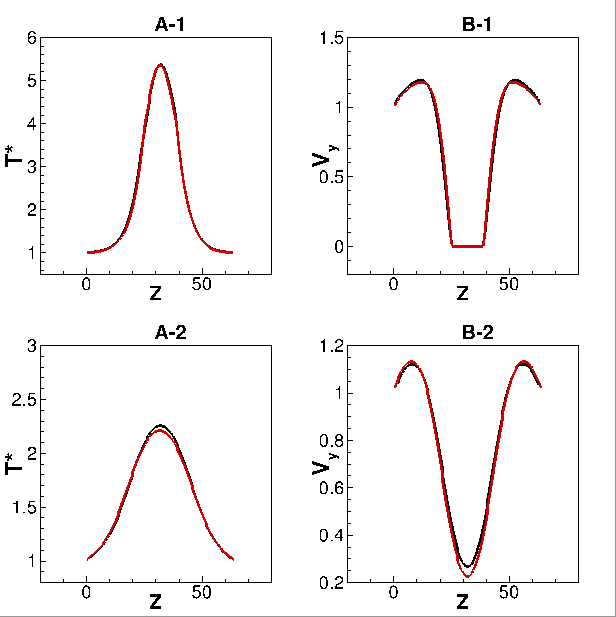}}% Images in 100% size
  \caption{Comparison of normalized temperature (A) and \textit{y}-velocity (B) between ML-solver (red) and Ansys Fluent (black) at \textit{y} = 21 (1) and \textit{y} = 40 (2) for a solid cube of size $13$m}
\label{fig:20}
\end{figure}

Next, we use the inferencing algorithm described in Figure \ref{fig:4} to predict the temperature and velocity fields for a cube of size, $13$ m, which has not been seen during training of the ML-solver. It may be observed from the contour plots in Figure \ref{fig:19} that the predictions of the ML-solver match well with those from Ansys Fluent \cite{fluent}. Additionally, it may also be observed that the peak normalized temperature in the solid domain is lower in this case as compared to when the size of the solid cube is smaller, attributing to the difference in their surfaces areas, thereby validating that a larger surface area results in an increased heat loss from the solid into the fluid domain. The line plot comparisons are shown in Figure \ref{fig:20}. It may be observed that the there is great agreement in the predictions of ML-solver as compared to Ansys Fluent \cite{fluent} and that the relative error in the predictions is less than $1$\%.  

\section{Conclusion}\label{sec:conclusion}
In this work, we have presented a novel ML-Solver, which uses important characteristics from existing PDE solvers for solving the system of steady, incompressible Navier-Stokes equation. The ML-solver does not require any training data and instead, generates and learns the PDE solutions simultaneously, during the training process. It uses discretization techniques to approximate the PDE residual at each voxel of a given computational domain and uses the \textit{L}-2 norm of the residual to update  network weights. The discretization schemes are implemented inside the computational graph to enable vectorization on GPU and provide access to numerous higher order and advanced numerical schemes that can enhance the accuracy as well as improve stability of the ML-solver, through physics-based regularization. In this work, we have extended the discretizations to unstructured domains by employing stair-step discretizations to provide flexibility in modeling different types of geometries as well as widely varying boundary conditions.

From the network architecture perspective, we introduce the DiscretizationNet, which is a generative CNN-based encoder-decoder network conditioned on geometry and boundary conditions. Separate autoencoders are constructed to learn lower-dimensional vectors (or encodings) for different geometry and boundary conditions. These encodings are used to enrich and parameterize the solution latent vector space of the generative network and thus allow for simultaneously generating and learning a wide range of solutions at different conditions in the same training session. Moreover, we employ a novel iterative capability in the network to mimic existing PDE solvers. In this implementation, the inputs to the generative model are replaced with outputs during network training, as the network learns to generate better solutions. This strategy is unique and we have observed that it provides better stability and faster convergence in comparison to other ML strategies, especially in cases when the ground truth solutions are not known. Additionally, we have proposed an algorithm for inferencing using the DiscretizationNet. The algorithm functions in the latent space to iteratively infer solutions using the trained model weights.

We have validated the ML-solver by solving the 3-D steady,incompressible Navier-Stokes equations on three different cases, (i) lid-driven cavity, (ii) laminar flow past a cylinder and (iii) conjugate heat transfer. Contour and line plot comparisons made with ANSYS Fluent R19.3 \citep{fluent} in all three cases show a good agreement. Additionally, it has been observed that the training for a large number of PDE solutions results in a stable convergence within $3$x$10^4$ training epochs.

The ML-solver proposed here can be extended to solve unsteady problems using LSTM-type networks \citep{hochreiter1997long}. The deficiencies in stair-step discretization, in computing accurate solutions near the boundaries can be mitigated by using a cut-cell of unstructured grid discretization. Moreover, the ML-solver in can be applied to other PDEs with complex physics as well as to develop computationally inexpensive low-dimensional models.

\bibliographystyle{elsarticle-num}
\bibliography{references}

\begin{thebibliography}{10}
\expandafter\ifx\csname url\endcsname\relax
  \def\url#1{\texttt{#1}}\fi
\expandafter\ifx\csname urlprefix\endcsname\relax\def\urlprefix{URL }\fi
\expandafter\ifx\csname href\endcsname\relax
  \def\href#1#2{#2} \def\path#1{#1}\fi

\bibitem{lee1990neural}
H.~Lee, I.~S. Kang, Neural algorithm for solving differential equations,
  Journal of Computational Physics 91~(1) (1990) 110--131.

\bibitem{lagaris1998artificial}
I.~E. Lagaris, A.~Likas, D.~I. Fotiadis, Artificial neural networks for solving
  ordinary and partial differential equations, IEEE transactions on neural
  networks 9~(5) (1998) 987--1000.

\bibitem{raissi2018hidden}
M.~Raissi, G.~E. Karniadakis, Hidden physics models: Machine learning of
  nonlinear partial differential equations, Journal of Computational Physics
  357 (2018) 125--141.

\bibitem{raissi2017physics}
M.~Raissi, P.~Perdikaris, G.~E. Karniadakis, Physics informed deep learning
  (part i): Data-driven solutions of nonlinear partial differential equations,
  arXiv preprint arXiv:1711.10561 (2017).

\bibitem{baydin2017automatic}
A.~G. Baydin, B.~A. Pearlmutter, A.~A. Radul, J.~M. Siskind, Automatic
  differentiation in machine learning: a survey, The Journal of Machine
  Learning Research 18~(1) (2017) 5595--5637.

\bibitem{dwivedi2019distributed}
V.~Dwivedi, N.~Parashar, B.~Srinivasan, Distributed physics informed neural
  network for data-efficient solution to partial differential equations, arXiv
  preprint arXiv:1907.08967 (2019).

\bibitem{sun2020surrogate}
L.~Sun, H.~Gao, S.~Pan, J.-X. Wang, Surrogate modeling for fluid flows based on
  physics-constrained deep learning without simulation data, Computer Methods
  in Applied Mechanics and Engineering 361 (2020) 112732.

\bibitem{zhu2019physics}
Y.~Zhu, N.~Zabaras, P.-S. Koutsourelakis, P.~Perdikaris, Physics-constrained
  deep learning for high-dimensional surrogate modeling and uncertainty
  quantification without labeled data, Journal of Computational Physics 394
  (2019) 56--81.

\bibitem{rao2020physics}
C.~Rao, H.~Sun, Y.~Liu, Physics-informed deep learning for incompressible
  laminar flows, arXiv preprint arXiv:2002.10558 (2020).

\bibitem{jin2020nsfnets}
X.~Jin, S.~Cai, H.~Li, G.~E. Karniadakis, Nsfnets (navier-stokes flow nets):
  Physics-informed neural networks for the incompressible navier-stokes
  equations, arXiv preprint arXiv:2003.06496 (2020).

\bibitem{zhuang2020learned}
J.~Zhuang, D.~Kochkov, Y.~Bar-Sinai, M.~P. Brenner, S.~Hoyer, Learned
  discretizations for passive scalar advection in a 2-d turbulent flow, arXiv
  preprint arXiv:2004.05477 (2020).

\bibitem{bar2018data}
Y.~Bar-Sinai, S.~Hoyer, J.~Hickey, M.~P. Brenner, Data-driven discretization:
  machine learning for coarse graining of partial differential equations,
  Preprint (2018).

\bibitem{hsieh2019learning}
J.-T. Hsieh, S.~Zhao, S.~Eismann, L.~Mirabella, S.~Ermon, Learning neural pde
  solvers with convergence guarantees, arXiv preprint arXiv:1906.01200 (2019).

\bibitem{stevens2020finitenet}
B.~Stevens, T.~Colonius, Finitenet: A fully convolutional lstm network
  architecture for time-dependent partial differential equations, arXiv
  preprint arXiv:2002.03014 (2020).

\bibitem{osher1988fronts}
S.~Osher, J.~A. Sethian, Fronts propagating with curvature-dependent speed:
  algorithms based on hamilton-jacobi formulations, Journal of computational
  physics 79~(1) (1988) 12--49.

\bibitem{rhie1983numerical}
C.~Rhie, W.~L. Chow, Numerical study of the turbulent flow past an airfoil with
  trailing edge separation, AIAA journal 21~(11) (1983) 1525--1532.

\bibitem{patankar1981calculation}
S.~V. Patankar, A calculation procedure for two-dimensional elliptic
  situations, Numerical heat transfer 4~(4) (1981) 409--425.

\bibitem{seo2011sharp}
J.~H. Seo, R.~Mittal, A sharp-interface immersed boundary method with improved
  mass conservation and reduced spurious pressure oscillations, Journal of
  computational physics 230~(19) (2011) 7347--7363.

\bibitem{tucker2000cartesian}
P.~Tucker, Z.~Pan, A cartesian cut cell method for incompressible viscous flow,
  Applied Mathematical Modelling 24~(8-9) (2000) 591--606.

\bibitem{chollet2015keras}
F.~Chollet, et~al., keras (2015).

\bibitem{fluent}
A.~Fluent, 19.3, theory guide, ansys (2019).

\bibitem{gelfgat2019linear}
A.~Y. Gelfgat, Linear instability of the lid-driven flow in a cubic cavity,
  Theoretical and Computational Fluid Dynamics 33~(1) (2019) 59--82.

\bibitem{kipf2016semi}
T.~N. Kipf, M.~Welling, Semi-supervised classification with graph convolutional
  networks, arXiv preprint arXiv:1609.02907 (2016).

\bibitem{hanocka2019meshcnn}
R.~Hanocka, A.~Hertz, N.~Fish, R.~Giryes, S.~Fleishman, D.~Cohen-Or, Meshcnn: A
  network with an edge, ACM Transactions on Graphics (TOG) 38~(4) (2019) 90.

\bibitem{hochreiter1997long}
S.~Hochreiter, J.~Schmidhuber, Long short-term memory, Neural computation 9~(8)
  (1997) 1735--1780.

\end{thebibliography}

\end{document}